\documentclass{llncs}
\usepackage{article}

\title{Writing Reusable Digital Topology Algorithms in a Generic Image
  Processing Framework}
\author{Roland Levillain\inst{1,2}, Thierry G\'eraud\inst{1,2},
  Laurent Najman\inst{2}}
\institute{EPITA Research and Development Laboratory (LRDE)\\
  14-16, rue Voltaire, FR-94276 Le Kremlin-Bic\^etre Cedex, France
  \and
  Universit\'e Paris-Est, Laboratoire d'Informatique Gaspard-Monge,
  \'Equipe A3SI, ESIEE Paris, Cit\'e Descartes, BP 99, FR-93162
  Noisy-le-Grand Cedex, France%
  \\
  \email{\{roland.levillain,thierry.geraud\}@lrde.epita.fr,
    l.najman@esiee.fr}%
}
\newcommand{\keywords}{Generic Programming, Interface, Skeleton, Complex}

\makeatletter
\hypersetup{pdftitle=\@title,
  pdfauthor={Roland Levillain, Thierry G\'eraud, Laurent Najman},
  pdfkeywords={\keywords}}
\makeatother

\begin{document}

\maketitle
\setcounter{footnote}{0}

\begin{abstract}
  Digital Topology software should reflect the generality of the
  underlying mathematics: mapping the latter to the former requires
  genericity.
  By designing generic solutions, one can effectively reuse digital
  topology data structures and algorithms.
  We propose an image processing framework focused on the Generic
  Programming paradigm in which an algorithm on the paper can be turned
  into a single code, written once and usable with various input
  types.
  This approach enables users to design and implement new methods at a
  lower cost, try cross-domain experiments and help generalize
  results.

\end{abstract}

\section{Introduction}


Like \ac{mm}, \ac{dt} has many applications in image analysis and
processing.  Both present sound mathematical foundations to handle
many types of discrete images.  In fact most methods from \acl{mm} or
\acl{dt} are not tied to a specific context
(image type, neighborhood, topology): they are most often described in
abstract and general terms.  Thus they are not limiting their field of
application.
However, software packages for \ac{mm} and \ac{dt} rarely take
(enough) advantage of this generality: an algorithm is sometimes
reimplemented for each image and/or each value type, or worse, written for
a unique input type.
Such implementations are not reusable because of
their lack of \emph{genericity}.  These limitations often come from
the implementation framework, which prohibits a generic design of
algorithms.  A recent and notable exception is the DGtal project,
which proposes \ac{dg} software tools and algorithms built in a generic
\Cxx framework \cite{dgtal.www.local}.

Thanks to the \ac{gp} paradigm, provided in particular by the \Cxx
language, one can design and implement generic frameworks.  This
paradigm is especially well-suited to the field of scientific
applications where the efficiency, widespread availability and
standardization of \Cxx are real assets.  To this end, we have
designed a paradigm dedicated to generic and efficient scientific
software \cite{geraud.08.mpool} and applied the idea of generic
algorithms to \ac{mm} in \ac{ip} \cite{levillain.09.ismm}, as
suggested by d'Ornellas and van den Boomgaard
\cite{dornellas.03.ijprai}.  The result of our experiments is a
generic library, Milena, part of the Olena image processing platform
\cite{olena.www}.

Lamy suggests to implement digital topology in \ac{ip} libraries
\cite{lamy.07.cmpb}.  The proposed solution, applied to the ITK
library \cite{ibanez.05.book,itk.www} ``\emph{works for any image dimension}''.
In this paper, we present a framework for the generic implementation
of \ac{dt} methods within the Milena library, working for
\emph{any image type} supporting the required notions (value types,
geometric and topological properties, etc.).  Such a generic framework
requires the definition
of concepts from the domain (in particular, of an image) to organize
data structures and algorithms, as explained in
\autoref{sec:genericity}.  Given
these concepts it is possible to write generic algorithms, like a
homotopic thinning operator making use of various definitions of the notion
of simple point.  We present a generic definition of such an operator
in \autoref{sec:implementation} and show some illustrations in
\autoref{sec:illustrations}.
\manualref{Section}{sec:conclusion} concludes on the extensibility of
this work
along different axes: existing algorithms, new data structures and
efficiency.

\section{Genericity in Image Processing}
\label{sec:genericity}


In order to design a generic framework for image processing, we have
previously proposed the following definition of an image
\cite{levillain.09.ismm}.

\begin{definition*}
An image $I$ is a function from a domain \Dom{} to a set of values
\Val{}; the elements of \Dom{} are called the \emph{sites} of $I$,
while the elements of \Val{} are its \emph{values}.  
\end{definition*}

For the sake of generality, we use the term \emph{site} instead of
\emph{point}; e.g. a site could represent a triangle of a surface mesh
used as the domain of an image.
Classical site sets used as image domains encompass hyperrectangles
(boxes) on regular $n$-dimensional grids, graphs and complexes (see
\autoref{sec:implementation}).

In the \ac{gp} paradigm, these essential notions (image, site set,
site, value) must be translated into interfaces called \emph{concepts}
in Milena (\concept{Image}, \concept{Site\_Set}, etc.)
\cite{levillain.10.icip}.  These interfaces contain the list of
\emph{services} provided by each type belonging to the concept, as
well as its \emph{associated types}.  For instance, a type satisfying
the \concept{Image} concept must provide a \code{domain()} routine (to
retrieve \Dom{}), as well as a \code{domain\_t} type (i.e. the type of
\Dom) satisfying the \concept{Site\_Set} concept.
Concepts act as contracts between providers (types
satisfying the concept) and users (algorithms expressing
requirements on their inputs and outputs through concepts).
For instance, the \code{breadth\_first\_thinning} routine from
\autoref{lst:breadth-first-thinning-cxx} expects the type \code{I} (of
the \code{input} image) to fulfill the requirements of the
\concept{Image} concept.  Likewise \code{nbh} must be a
\concept{Neighborhood}; and \code{is\_simple} and \code{constraint}
must be functions taking a value of arbitrary type and returning a
Boolean value (\code{Function\_v2b} concept).

\section{Generic Implementation of Digital Topology}
\label{sec:implementation}

Let us consider the example of homotopic skeletonization by thinning.
Such an operation can be obtained by the removal of \emph{simple
  points} (or simple sites in the Milena parlance) using
\autoref{lst:breadth-first-thinning} \cite{bertrand.07.chapter}.  A
point of an object is said to be simple if its deletion does not
change the topology of the object.  This algorithm takes an object $X$
and a constraint $K$ (a set of points that must not be removed) and
iteratively deletes simple points of $X \backslash K$ until stability
is reached.
\autoref{lst:breadth-first-thinning} is an example of an algorithm
with a general definition that could be applied to many input types in
theory.  But in practice, software tools often allow a limited set of
such input types (sometimes just a single one), because some operations
(like ``is simple'') are tied to the definition of the algorithm
\cite{levillain.09.ismm}.

\autoref{lst:breadth-first-thinning-gen} shows a more general version
of \autoref{lst:breadth-first-thinning}, where im\-ple\-men\-ta\-tion-specific
elements have been replaced by \emph{mutable} parts:
a predicate stating whether a point $p$ is simple with respect to a
set $X$ (\emph{is\_simple});
a routine ``detaching'' a (simple) point $p$ from a set $X$
(\emph{detach});
and a predicate declaring whether a condition (or a set of conditions)
on $p$ is satisfied before considering it for removal
(\emph{constraint}).
The algorithm takes these three functions as arguments in addition to
the input $X$.
\autoref{lst:breadth-first-thinning-gen} is a good candidate for a
generic \Cxx implementation of the breadth-first thinning strategy and
has been implemented as \autoref{lst:breadth-first-thinning-cxx} in
Milena\footnote{In \autoref{lst:breadth-first-thinning-cxx},
  \code{mln\_ch\_value(I, V)} and \code{mln\_concrete(I)} are helper
  macros.  The former returns the image type associated to \code{I}
  where the value type has been set to \code{V}.  The latter returns
  an image type corresponding to \code{I} with actual data storage
  capabilities.  In many cases, \code{mln\_concrete(I)} is simply
  equal to \code{I}.}.  This algorithms implements the breadth-first
traversal by using a \acr{fifo} queue.  The set $X$ is represented by
a binary image ($\Val = \{\mbox{true}, \mbox{false}\}$), that must be
compatible with operations performed within the algorithm.  Inputs
\emph{is\_simple}, \emph{detach} and \emph{constraint}\footnote{Note
  that the notion of ``constraint'' is not the same in
  \autoref{lst:breadth-first-thinning} and
  \autoref{lst:breadth-first-thinning-cxx}: in the former, it is the
  set of points to preserve, while in the latter is it a predicate
  that a candidate point must pass to be removed.} have been turned
into function objects (also called \emph{functors}).  The
\code{breadth\_first\_thinning} routine creates and returns an image with type
\code{mln\_concrete(I)}; it is an image type equivalent to \code{I}
that allows to store data for every sites independently (which is not
the case for some image types).

\begin{lstlisting}[language=Algorithm,%
  float,%
  caption={Breadth-First Thinning.},%
  label={lst:breadth-first-thinning}]
Data : $E$ (a set of points/sites),
$X \subseteq E$ (initial set of points),
$K \subseteq X$ (a set of points (constraint) that cannot be removed)
Result : $X$
$P$ <- $\{ \, p \in X \;|\; p \mbox{ is simple for } X \, \}$
while $P \neq \emptyset$ do
  $S$ <- $\emptyset$
  for each $p \in P$ do
    if $p \not\in K$ and $p \mbox{ is simple for } X$ then
      $X$ <- $X - \{p\}$
      for each $n \,\in\, \mathcal{N}(p) \cap X$ do
        $S$ <- $S \cup \{ n \}$
  $P$ <- $\emptyset$
  for each $p \in S$ do
    if $p \mbox{ is simple for } X$ then $P$ <- $P \cup \{p\}$
\end{lstlisting}

\begin{lstlisting}[language=Algorithm,%
  escapechar={@},%
  emph={is_simple,detach,constraint},%
  float,%
  caption={A generic version of
    \autoref{lst:breadth-first-thinning}.},%
  label={lst:breadth-first-thinning-gen}]
Data : $E$, $X \subseteq E$, $\mathcal{N}$ (neighborhood),
is_simple (a function saying whether a point is simple),
detach (a routine detaching a point from $X$),
constraint (a function representing a @constraint@)
Result : $X$
$P$ <- $\{ \, p \in X \;|$ is_simple($p$, $X$) $\}$
while $P \neq \emptyset$ do
  $S$ <- $\emptyset$
  for each $p \in P$ do
    if constraint($p$) and is_simple($p$, $X$) then
      $X$ <- detach($X$, $p$)
      for each $n \,\in\, \mathcal{N}(p) \cap X$ do
        $S$ <- $S \cup \{ n \}$
  $P$ <- $\emptyset$
  for each $p \in S$ do
    if is_simple($p$, $X$) then $P$ <- $P \cup \{p\}$
\end{lstlisting}  

\lstinputlisting[language=C++,basicstyle={\ttfamily},%
emph={is_simple,detach,constraint},%
emph={[2]input_,nbh_,is_simple_,constraint_,%
  input,nbh,output,queue,in_queue,p,n},%
emphstyle={[2]\color{darkgreen}\bfseries},%
morekeywords={[1]for_all},%
morekeywords={[2]Image,I,Neighborhood,N,Function_v2b,F,G,H,%
  psite,p_queue_fast},%
float,%
texcl,%
caption={A generic \Cxx implementation of %
  \autoref{lst:breadth-first-thinning-gen} in Milena.
  Functors are highlighted.},%
label={lst:breadth-first-thinning-cxx}]%
{breadth-first-thinning.cc}

\subsection*{Simple Point Characterization Implementation}

There are local characterizations of simple points in 2D, 3D and 4D,
which can lead to \ac{lut} based implementations
\cite{couprie.09.pami}.  However, since the number of configurations
of simple and non-simple points in $\ZZ^d$ is $2^{3^d - 1}$,
this approach can only be used in practice in 2D (256 configurations,
requiring a \ac{lut} of 32 bytes) and possibly in 3D (67,108,864
configurations, requiring a \ac{lut} of 8 megabytes).  The 4D case exhibits
$2^{80}$ configurations, which is intractable using a \ac{lut}, as it would
need 128 zettabytes (128 billions of terabytes) of memory.
Couprie and Bertrand have proposed a more general framework for
checking for simple points using cell complexes
\cite{couprie.09.pami} and the collapse operation.
Intuitively, complexes can be seen as a generalization of graphs.  An
informal definition of a \emph{simplicial complex} (or simplicial
$d$-complex) is ``a set of simplices'' (plural of simplex), where a
simplex or $n$-simplex is the simplest manifold that can be created
using $n$ points (with $0 \leq n \leq d$).  A 0-simplex is a point, a
1-simplex a line segment, a 2-simplex a triangle, a 3-simplex a
tetrahedron.  A graph is indeed a 1-complex.
\manualref{Figure}{fig:complex:simplicial} shows an example of a
simplicial complex.
Likewise, a \emph{cubical complex} or cubical $d$-complex can be
thought as a set of $n$-faces (with $0 \leq n \leq d$) in $\ZZ^d$,
like points (0-faces), edges (1-faces), squares (2-faces), cubes
(3-faces) or hypercubes (4-faces).
\manualref{Figure}{fig:complex:cubical} depicts a cubical complex sample.

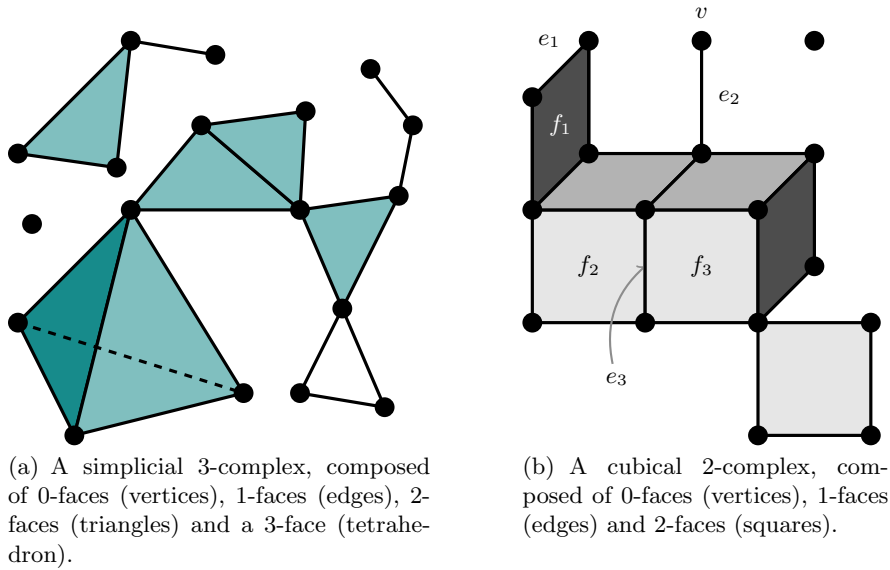
\begin{figure}[tbp]
  \centering
  \tikzstyle{edge}=[very thick]
  \subfigure[A simplicial 3-complex, composed of 0-faces (vertices),
  1-faces (edges), 2-faces (triangles) and a 3-face (tetrahedron).]{%
    \begin{tikzpicture}[scale=0.75]
      \tikzstyle{hidden edge}=[very thick, dashed]
      \tikzstyle{triangle}=[fill={rgb:green,5;blue,5;white,10}, draw=black,
      very thick]
      \tikzstyle{dark triangle}=[fill={rgb:green,5;blue,5;white,1}, draw=black,
      very thick]
      \filldraw[dark triangle] (1, 0) -- (2, 4) -- (0, 2) -- cycle;
      \filldraw[triangle] (1, 0) -- (4, 0.75) -- (2, 4) -- cycle;
      \draw[hidden edge] (0, 2) -- (4, 0.75);
      \filldraw[triangle] (2, 4) -- (5, 4) -- (3.25, 5.5) -- cycle;
      \filldraw[triangle] (5, 4) -- (3.25, 5.5) -- (5.1, 5.75) -- cycle;
      \filldraw[triangle] (5, 4) -- (5.75, 2.25) -- (6.75, 4.25) -- cycle;
      \draw[edge] (5.75, 2.25) -- (5, 0.75) -- (6.5, 0.5) -- cycle;
      \draw[edge] (6.75, 4.25) -- (7, 5.5) -- (6.25, 6.5);
      \filldraw[triangle] (0, 5) -- (1.75, 4.75) -- (2, 7) -- cycle;
      \draw[edge] (2, 7) -- (3.5, 6.75);
      \foreach \p in {%
        (0, 2), (0.25, 3.75), (0, 5), (1, 0), (1.75, 4.75), (2, 4),%
        (2, 7), (3.25, 5.5), (3.5, 6.75), (4, 0.75), (5, 0.75),%
        (5, 4), (5.1, 5.75), (5.75, 2.25), (6.75, 4.25), (6.5, 0.5),%
        (7, 5.5), (6.25, 6.5)%
      }
      \fill \p circle (5pt);
    \end{tikzpicture}
    \label{fig:complex:simplicial}%
  }
  \hspace{1cm}
  \subfigure[A cubical 2-complex, composed of 0-faces (vertices),
  1-faces (edges) and 2-faces (squares).]{%
    \begin{tikzpicture}[scale=0.75]
      \tikzstyle{front quad}=[fill=black!10!white, draw=black, very thick]
      \tikzstyle{top quad}=[fill=black!30!white, draw=black, very thick]
      \tikzstyle{right quad}=[fill=black!70!white, draw=black, very thick]
      pattern color=gray!75, pattern={north west lines}]
      \filldraw[front quad] (0, 0) rectangle (2, 2);
      \node at (1, 1) {$f_2$};
      \node (e3label) at (1.5, -1) {$e_3$};
      \path[->] (e3label) edge [color=gray!90, thick, bend left] (2, 1);
      \filldraw[front quad] (2, 0) rectangle (4, 2);
      \node at (3, 1) {$f_3$};
      \fill[right quad] (0, 2) -- (0, 4) -- (1, 5) -- (1, 3) -- cycle;
      \node at (0.5, 3.5) [text=white] {$f_1$};
      \draw[edge] (0, 2) -- (0, 4);
      \draw[edge] (1, 3) -- (1, 5);
      \node at (0.3, 5) {$e_1$};
      \filldraw[top quad] (0, 2) -- (2, 2) -- (3, 3) -- (1, 3) -- cycle;
      \filldraw[top quad] (2, 2) -- (4, 2) -- (5, 3) -- (3, 3) -- cycle;
      \filldraw[right quad] (4, 0) -- (4, 2) -- (5, 3) -- (5, 1) -- cycle;
      \filldraw[front quad] (4, 0) rectangle (6, -2);
      \draw[edge] (3, 3) -- (3, 5);
      \node at (3.5, 4) {$e_2$};
      \foreach \p in {%
        (1, 5), (5, 5), (3, 5),%
        (0, 4),%
        (1, 3), (3, 3), (5, 3),%
        (0, 2), (2, 2), (4, 2),%
        (5, 1),%
        (0, 0), (2, 0), (4, 0), (6, 0),%
        (4, -2), (6, -2)%
      } \fill \p circle (5pt);
      \node at (3, 5.5) {$v$};
    \end{tikzpicture}
    \label{fig:complex:cubical}%
  }
  \caption{Examples of cell complexes.}
  \label{fig:complex}
\end{figure}

Complexes support a topology-preserving transformation called
\emph{collapse}.
An \emph{elementary collapse} removes a free pair of faces of a
complex, like the square face $f_1$ and its top edge $e_1$, or the edge
$e_2$ and its top vertex $v$, in \autoref{fig:complex:cubical}.  The
pair $(f_2, e_3)$ cannot be removed, since $e_3$ also belongs to $f_3$.
Successive elementary collapses form a \emph{collapse sequence} that
can be used to remove simple points.  Collapse-based implementations
of simple-point deletion can always be used in 2D, 3D and 4D,
though they are less efficient than their \ac{lut}-based counterparts.  On
the other hand, they provide some genericity as the collapse operation
can have a single generic implementation on complexes regardless of
their structure.

\section{Illustrations}
\label{sec:illustrations}

\newcommand{\myww}{0.45\textwidth}
\newcommand{\mywww}{0.29\textwidth}

Using this generic approach, \autoref{lst:breadth-first-thinning-cxx}
can be used to compute skeletons of various input images.


\subsection{Skeleton of a 2D Binary Image}

%
%

\begin{figure}[tbp]
  \centering
  \subfigure[2D binary image.]{
    \fbox{\includegraphics[width=\mywww]{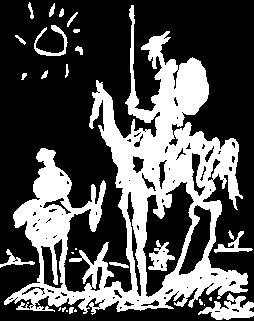}}
    \label{fig:picasso}
  }
  \subfigure[Skeleton of \subref{fig:picasso} with no constraint]{
    \fbox{\includegraphics[width=\mywww]{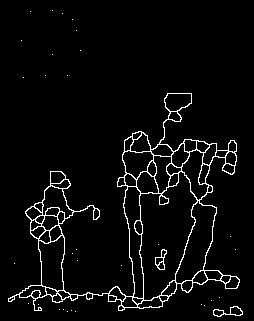}}
    \label{fig:picasso-skel-unconstrained}
  }
  \subfigure[Skeleton of \subref{fig:picasso} where end
  points of the initial image have been preserved.]{
    \fbox{\includegraphics[width=\mywww]{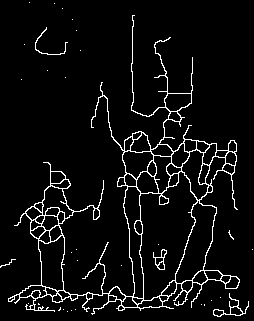}}
    \label{fig:picasso-skel-with-end-points}
  }
  \caption{Computation of skeletons from a 2D binary regular image.}
  \label{fig:picasso-skel}
\end{figure}

Our first illustration uses a classical 2D binary image built on a
square grid (\autoref{fig:picasso}).  The following lines produces the
result shown on \autoref{fig:picasso-skel-unconstrained}.
\begin{lstlisting}[language=C++,basicstyle={\ttfamily},frame={tb},%
% Identifiers.
emph={[2]input,output},%
emphstyle={[2]\color{darkgreen}\bfseries},%
% More types.
morekeywords={[2]image2d,neighb2d,I,N,%
  is_simple_point2d,detach_point,no_constraint}]
typedef image2d<bool> I;
typedef neighb2d N;
I output =
  breadth_first_thinning(input,
                         c4(),
                         is_simple_point2d<I, N>(c4(), c8()),
                         detach_point<I>(),
                         no_constraint());
\end{lstlisting}
\code{I} and \code{N} are introduced as aliases of the image and
neighborhood types for convenience.  The
\code{breadth\_first\_thinning} algorithm is called with five
arguments, as expected.  The first two ones are the input image and
the (4-connectivity) neighborhood used in the algorithm.  The last
three ones are the functors governing the behavior of the thinning
operator.  The call \code{is\_simple\_point2d<I, N>(c4(), c8())}
creates a simple point predicate based on the computation of the 2D
connectivity numbers \cite{bertrand.07.chapter} associated with the
4-connectivity for the foreground and the 8-connectivity for the
background.  To compute these numbers efficiently,
\code{is\_simple\_point2d} uses a \ac{lut} containing all the possible
configurations in the 8-connectivity neighborhood of a pixel.
\code{detach\_point<I>} is a simple functor removing a pixel by giving
it the value ``false''.  Finally, \code{no\_constraint} is an empty
functor representing a lack of constraint.

We also present a variation of the previous example where the fifth
argument passed to the function is an actual constraint, preserving
all end points of the initial image (see
\autoref{fig:picasso-skel-with-end-points}).  This result is obtained
by invoking the generic functor \code{is\_not\_end\_point} in the
following lines.  This call creates a predicate characterizing end
points by counting their number of neighbors.

\begin{lstlisting}[language=C++,basicstyle={\ttfamily},frame={tb},%
% Emphasized terms.
emph={is_not_end_point},%
% Identifiers.
emph={[2]input,output_with_end_points},%
emphstyle={[2]\color{darkgreen}\bfseries},%
% More types.
morekeywords={[2]image2d,neighb2d,I,N,%
  is_simple_point2d,detach_point}]
I output_with_end_points =
  breadth_first_thinning(input,
                         c4(),
                         is_simple_point2d<I, N>(c4(), c8()),
                         detach_point<I>(),
                         is_not_end_point<I, N>(c4(), input));
\end{lstlisting}

\subsection{Skeleton of a 3D Binary Image}

%
%

\begin{figure}[tbp]
  \centering
  \subfigure[3D binary image.]{
    \fbox{\includegraphics[width=\myww]{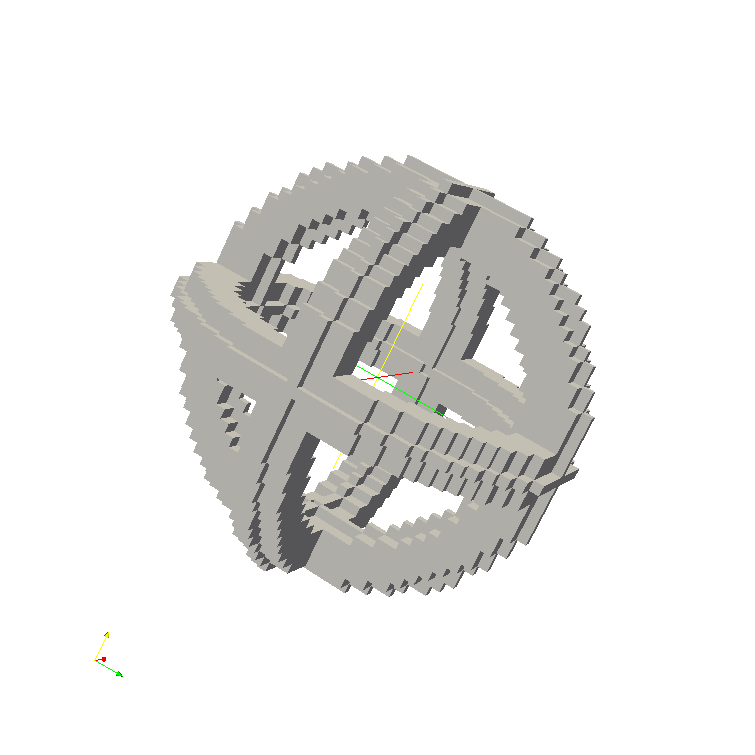}}
    \label{fig:image3d-skel-unconstrained-input}
  }
  \subfigure[Skeleton (dark voxels) of \subref{fig:picasso} with no
  constraint superimposed on the initial image (light voxels).]{
    \fbox{\includegraphics[width=\myww]{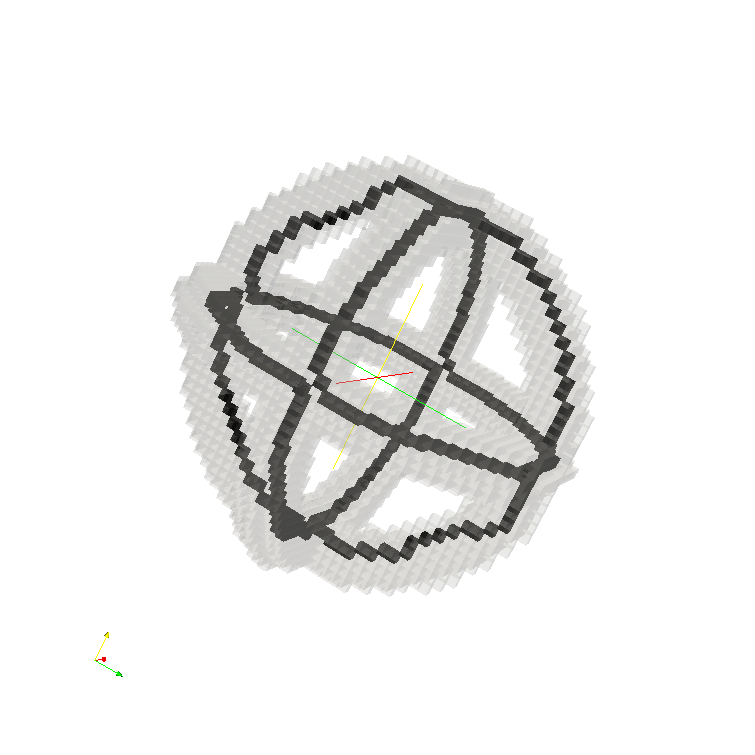}}
    \label{fig:image3d-skel-unconstrained-input+skel}
  }
  \caption{Computation of the skeleton of 3D binary regular image.}
  \label{fig:image3d-skel-unconstrained}
\end{figure}

This second example in 3D is similar to the previous one in 2D.  The
domain of the image is a box on a cubical grid;
the 26- and the 6-connectivity are respectively used for the
foreground and the background.  The output of
\autoref{fig:image3d-skel-unconstrained-input+skel} is obtained from
the 3D volume shown in \autoref{fig:image3d-skel-unconstrained-input}
with the following lines.
\begin{lstlisting}[language=C++,basicstyle={\ttfamily},frame={tb},%
% Identifiers.
emph={[2]input,output},%
emphstyle={[2]\color{darkgreen}\bfseries},%
% More types.
morekeywords={[2]image3d,neighb3d,I,N,%
  is_simple_point3d,detach_point,no_constraint}]
typedef image3d<bool> I;
typedef neighb3d N;
I output =
  breadth_first_thinning(input,
                         c26(),
                         is_simple_point3d<I, N>(c26(), c6()),
                         detach_point<I>(),
                         no_constraint());
\end{lstlisting}
The only real difference with the previous example is the use of the
functor \code{is\_simple\_point3d}.  The default implementation of
this predicate uses an on-the-fly computation of 3D connectivity
numbers.  We have also implemented a version based on a precomputed
\ac{lut} which showed significant speed-up improvements.

Please note that the predicates \code{is\_simple\_point2d} and
\code{is\_simple\_point3d} are specifically defined for a given
topology in order to preserve performances.

\subsection{Thick Skeleton of a 3D Mesh Surface}

%
%

\begin{figure}[tbp]
  \centering
  \subfigure[Triangle mesh surface.]{
    \fbox{\includegraphics[width=\mywww]{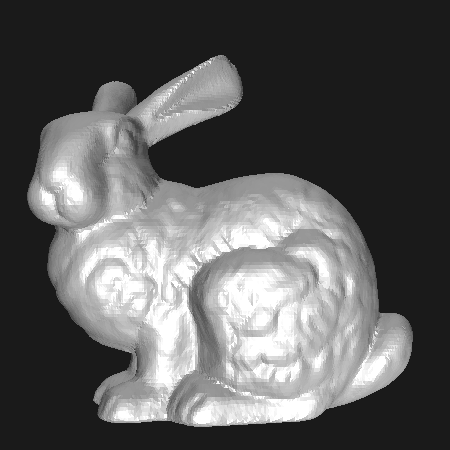}}
    \label{fig:bunny-holefilled-pinv-curv-skel:mesh}
  }
  \subfigure[Surface curvature.]{
    \fbox{\includegraphics[width=\mywww]{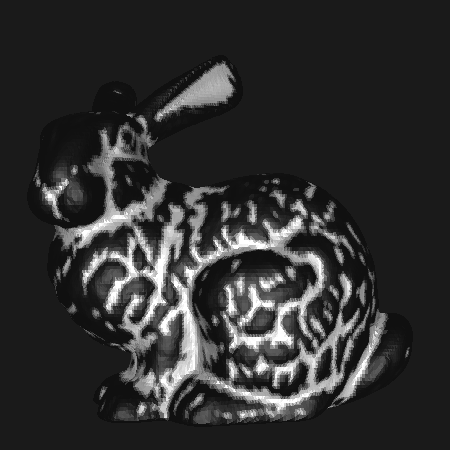}}
    \label{fig:bunny-holefilled-pinv-curv-skel:curv}
  }
  \subfigure[Surface skeleton.]{
    \fbox{\includegraphics[width=\mywww]{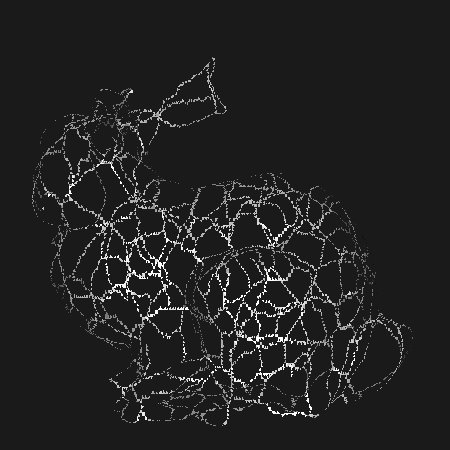}}
    \label{fig:bunny-holefilled-pinv-curv-skel:skel}
  }
  \caption{Computation of a skeleton using breadth-first thinning.
    The triangle mesh surface
    \ref{fig:bunny-holefilled-pinv-curv-skel:mesh}
    (35,286 vertices and 70,568 triangles)
    is seen as a simplicial 2-complex.  The image of
    curvature \ref{fig:bunny-holefilled-pinv-curv-skel:curv} is
    computed on the edges of the mesh, and simplified using
    an area opening filter.  All curvature regional minima are then
    removed from the mesh, and the skeleton
    \ref{fig:bunny-holefilled-pinv-curv-skel:skel} is obtained with
    \autoref{lst:breadth-first-thinning-cxx} using the collapse
    operation.}
  \label{fig:bunny-holefilled-pinv-curv-skel}
\end{figure}


In this third example, we manipulate discrete mesh surfaces composed
of triangles.  The input of the thinning operator is a surface
containing ``holes'', obtained from the mesh shown in
\autoref{fig:bunny-holefilled-pinv-curv-skel:mesh} by removing
triangles located in regional minima of the surface's curvature (darkest
areas of \autoref{fig:bunny-holefilled-pinv-curv-skel:curv}).  The
result presented in \autoref{fig:bunny-holefilled-pinv-curv-skel:skel}
is obtained with the following lines.  Types are not shown to make
this code more readable.
\begin{lstlisting}[language=C++, basicstyle={\ttfamily},frame={tb},%
% Identifiers.
emph={[2]input,nbh,is_simple_triangle,detach_triangle,output},%
emphstyle={[2]\color{darkgreen}\bfseries},%
% More types.
morekeywords={[2]no_constraint}]
output = breadth_first_thinning(input,
                                nbh,
                                is_simple_triangle,
                                detach_triangle,
                                no_constraint());
\end{lstlisting}
In the previous code, \code{input} is a triangle-mesh surface
represented by an image built on a simplicial 2-complex and \code{nbh}
represents an adjacency relationship between triangles sharing a
common edge.  Function objects \code{is\_simple\_triangle} and
\code{detach\_triangle} are operations compatible with \code{input}'s
type; they are generic routines based on the collapse operation
mentioned in \autoref{sec:implementation}, working with any complex-based
binary image.

The \code{input} image is constructed so that the sites browsed by the
\code{for\_all} loops in \autoref{lst:breadth-first-thinning-cxx} are
only 2-faces (triangles), while preserving access to values at 1-faces
and 0-faces.  Thus, even though they receive 2-faces as input
parameters, \code{is\_simple\_triangle} and \code{detach\_triangle}
are able to inspect the adjacent 1-faces and 0-faces and determine whether
and how a triangle can be completely detached from the surface through
a collapse sequence.

The resulting skeleton is said to be \emph{thick}, since it is
composed of triangles connected by a common edge.  The corresponding
complex is said to be pure, as it does not contain isolated 1-faces or
0-faces (that are not part of a 2-face).

\subsection{Thin Skeleton of a 3D Mesh Surface}

%
%

\begin{figure}[tbp]
  \centering
  \subfigure[Ultimate 2-collapse of
  \autoref{fig:bunny-holefilled-pinv-curv-skel:mesh}.]{
    \fbox{\includegraphics[width=\myww]{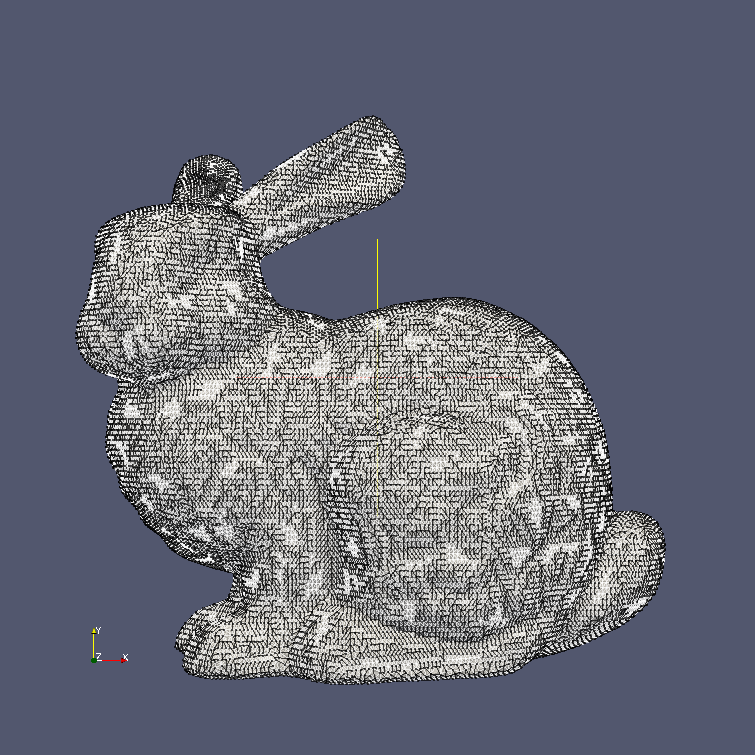}}
    \label{fig:bunny-holefilled-max-curv-2-collapse}
  }
  \subfigure[Ultimate 1-collapse of
  \subref{fig:bunny-holefilled-max-curv-2-collapse}]{
    \fbox{\includegraphics[width=\myww]{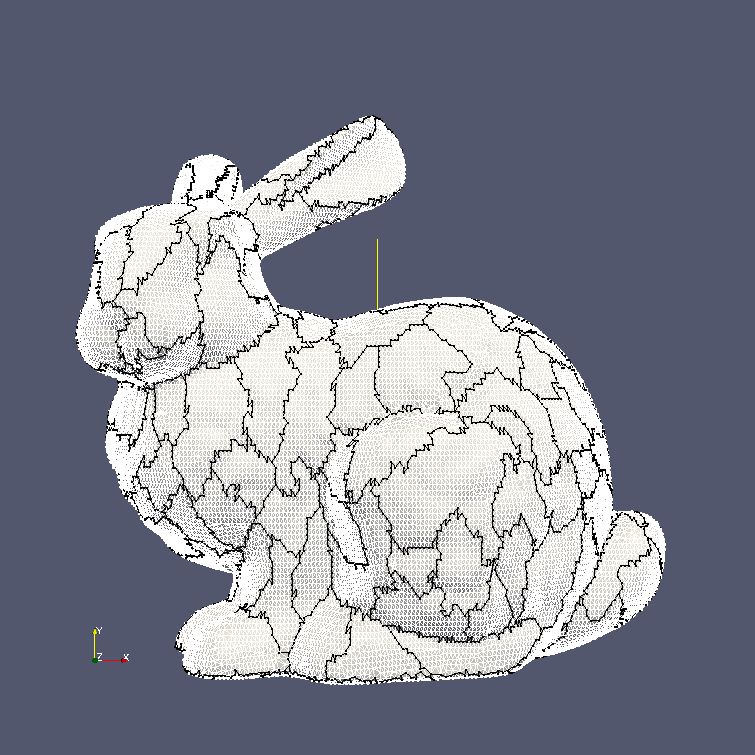}}
    \label{fig:bunny-holefilled-max-curv-1-collapse}
  }
  \caption{Thin skeleton obtain by 2- and 1-collapse.}
  \label{fig:bunny-holefilled-max-curv-collapses}
\end{figure}

To obtain a \emph{thin} skeleton, we can use a strategy based on
successive $n$-collapse operations, with $n$ decreasing
\cite{cousty.09.iwcia}.  From the input of the previous example, we
can obtain a ultimate 2-collapse by removing all simple pairs composed
of a 2-face and a 1-face (a triangle and an adjacent edge).  The following
lines compute such an ultimate 2-collapse.  The iteration on
\code{input}'s domain is still limited to triangles (2-faces).
\begin{lstlisting}[language=C++, basicstyle={\ttfamily},frame={tb},%
% Identifiers.
emph={[2]input,nbh,is_triangle_in_simple_pair,%
  detach_triangle_in_simple_pair,collapse2},%
emphstyle={[2]\color{darkgreen}\bfseries},%
% More types.
morekeywords={[2]no_constraint}]
collapse2 = breadth_first_thinning(input,
                                   nbh,
                                   is_triangle_in_simple_pair,
                                   detach_triangle_in_simple_pair,
                                   no_constraint());
\end{lstlisting}
Functor \code{is\_triangle\_in\_simple\_pair} checks whether a given
triangle is part of a simple pair, and if so
\code{detach\_triangle\_in\_simple\_pair} is used to remove the pair.
Thinning the initial surface with this ``simple site'' definition
produces a mesh free of 2-faces (triangles), as shown in
\autoref{fig:bunny-holefilled-max-curv-2-collapse}.

From this first skeleton, we can compute an ultimate 1-collapse, by
removing all simple pairs composed of an edge (1-face) and a vertex
(0-face).  This skeleton is produced with the following code, where
\code{input2} is an image created from \code{collapse2}, and for which
the domain of has been set to the edges of the complex, (instead of
the triangles).
\begin{lstlisting}[language=C++, basicstyle={\ttfamily},frame={tb},%
% Identifiers.
emph={[2]input2,nbh,is_edge_in_simple_pair,%
  detach_edge_in_simple_pair,collapse1},%
emphstyle={[2]\color{darkgreen}\bfseries},%
% More types.
morekeywords={[2]no_constraint}]
collapse1 = breadth_first_thinning(input2,
                                   nbh,
                                   is_edge_in_simple_pair,
                                   detach_edge_in_simple_pair,
                                   no_constraint());
\end{lstlisting}
Here \code{is\_edge\_in\_simple\_pair} and
\code{detach\_edge\_in\_simple\_pair} respectively test and remove an
edge along with a vertex that form a simple pair.  The result is a
simplified skeleton, with no isolated branches, as the lack of
constraint (\code{no\_constraint}) does not preserve them.  The output
of the ultimate 1-collapse on the bunny mesh is depicted in
\autoref{fig:bunny-holefilled-max-curv-1-collapse}.  It contains the
crest lines that form the boundaries of catchment basins, such as in
the watershed transform, and, in addition, the crest lines that make
the previous ones connect one to another.

Note that in both cases, the neighborhood object \code{nbh} is the
same, as it represents the adjacency of two $n$-faces connected by a
common adjacent $(n-1)$-face.  In the case of the 2-collapse, the
neighborhood of a site (triangle) is the set of adjacent triangles
connected by an edge, while in the case of the 1-collapse, the
neighborhood of a site (edge) is the set of adjacent edges
connected by a vertex.

\subsection{Execution Times}

\autoref{tab:times} shows the execution times of the previous
illustrations, computed on a PC running Debian GNU/Linux 6.0.4,
featuring an Intel Pentium~4 CPU running at 3.4~GHz with 2~GB RAM at
400~MHz, using the \Cxx compiler \command{g++} (GCC) version 4.4.5,
invoked with optimization option \samp{-03}.  The first three test
cases use a simple point criterion based on connectivity numbers,
while the last three use a collapse-based definition.

\begin{table}
  \centering
  \caption[Execution times of
  \autoref{lst:breadth-first-thinning-cxx} for various inputs.]{Execution
    times of \autoref{lst:breadth-first-thinning-cxx} for various
    inputs.  Figures correspond to the time spent in the
    \code{breadth\_first\_thinning} routine only.}
  \label{tab:times}
  \begin{tabular}{|l|l|l|l|r|}
    \hline
    Input & Input size & Constraint & Output & \multicolumn{1}{l|}{Time}\\
    \hline
    \hline
    2D image (\autoref{fig:picasso})
    & \multirow{2}{*}{321 $\times$ 254 pixels} 
    & None
    & \autoref{fig:picasso-skel-unconstrained}
    & 0.08 s\\
    \cline{1-1}\cline{3-5}
    2D image (\autoref{fig:picasso})
    &
    & End points
    & \autoref{fig:picasso-skel-with-end-points}
    & 0.10 s\\
    \hline
    3D image (\autoref{fig:image3d-skel-unconstrained-input})
    & 41 $\times$ 41 $\times$ 41 voxels 
    & None
    & \autoref{fig:image3d-skel-unconstrained-input+skel}
    & 2.67 s\\
    \hline
    Mesh (2-faces only)
    (\autoref{fig:bunny-holefilled-pinv-curv-skel:mesh})
    &
    & None
    & \autoref{fig:bunny-holefilled-pinv-curv-skel:skel}
    & 159.53 s\\
    \cline{1-1}\cline{3-5}
    Mesh (2- and 1-faces)
    & \hspace{5pt}35,286 0-faces +
    & \multirow{2}{*}{None}
    & \autoref{fig:bunny-holefilled-max-curv-2-collapse}
    & \multirow{2}{*}{68.78 s}\\
    (\autoref{fig:bunny-holefilled-pinv-curv-skel:mesh})
    & 105,852 1-faces +
    & 
    & (2-collapse)
    &
    \\
    \cline{1-1}\cline{3-5}
    Mesh (1- and 0-faces)
    & \hspace{5pt}70,568 2-faces
    & \multirow{2}{*}{None}
    & \autoref{fig:bunny-holefilled-max-curv-1-collapse}
    & \multirow{2}{*}{46.18 s}\\
    (\autoref{fig:bunny-holefilled-max-curv-2-collapse})
    &
    &
    & (1-collapse)
    &
    \\
    \hline
  \end{tabular}
\end{table}


\section{Conclusion}
\label{sec:conclusion}

We have presented building blocks to implement reusable \acl{dt}
algorithms in an \acl{ip} framework, Milena.  Given a set of
theoretical constraints on its inputs, an algorithm can be written
once and reused with many compatible image types.  This design has
previously been proposed for \acl{mm}, and can be applied to virtually
any image processing field.
Milena is Free Software released under the \acl{gpl}, and can be
freely downloaded from \url{http://olena.lrde.epita.fr/}.

\begin{figure}[tbp]
  \centering
  \subfigure[2D gray-level image.]{
    \fbox{\includegraphics[width=\myww]{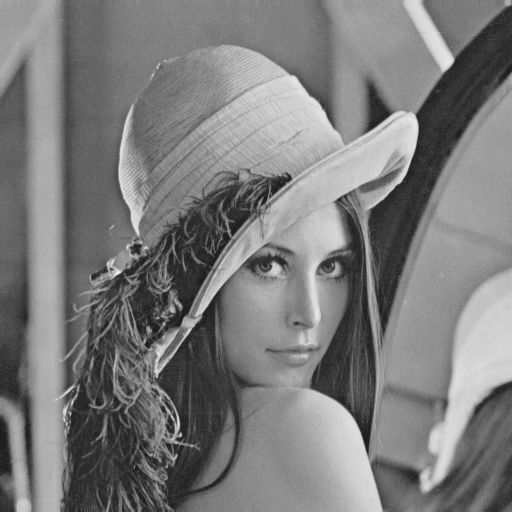}}
    \label{fig:lena}
  }
  \subfigure[Gray-level skeleton.]{
    \fbox{\includegraphics[width=\myww]{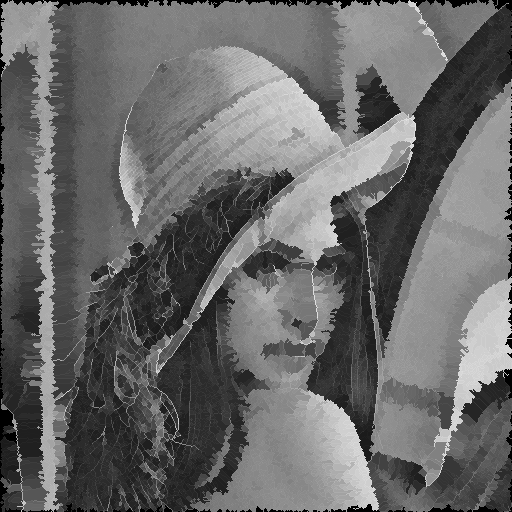}}
    \label{fig:lena-skel}
  }
  \caption{Computation of a gray-level skeleton.}
  \label{fig:gray-level-skel}
\end{figure}

A strength of generic designs is their ability to extend and scale
easily and efficiently.
First, generic algorithms are extensible because of their
parameterization.  For instance, the
behavior of \autoref{lst:breadth-first-thinning-cxx} can be
changed by acting on the simple point definition or the
set of constraints.
The scope of this algorithm, initially designed
to produce homotopic thinnings of binary skeleton, can even be
extended further to handle gray-level images and produce gray-level
thinnings.  From a theoretical point of view, gray-level images can be
processed by decomposing them
into different sections.  The equivalent of \emph{detaching} a
\emph{simple point} in a binary image is the \emph{lowering} of a
\emph{destructible point} in a gray-level context
\cite{couprie.01.jei}.
We have been able to produce gray-level skeletons with
\autoref{lst:breadth-first-thinning-cxx} by simply replacing the
\code{is\_simple} and \code{detach} operations by
\code{is\_destructible} and \code{lower} functors (see
\autoref{fig:gray-level-skel}).  In the case of a 2D regular images on
a square grid, this operation is straightforward as a destructible
point can also be characterized locally using new definitions of
connectivity numbers.

Generic algorithms can thereafter be turned into \emph{patterns} or
\emph{canvases} \cite{dornellas.01.phd} allowing the implementation of
many algorithms sharing a common core.  For example Milena implements
morphological algorithms like dilation and erosion, reconstructions,
etc. depending on the browsing strategy.  \acl{dt} could also benefit
from a canvas-based approach.
The framework can also be extended with respect to data
structures.  Milena provides site sets based on boxes, graphs and
complexes, but more can be added to the library (e.g. combinatorial
maps, orders, etc.) and benefit from existing algorithms and tools.

Finally, our approach can take advantage of \emph{properties} of input
types (regularity of the site set, isotropic adjacency relationship,
etc.) and allow users to write specialized versions of their
algorithms for such subsets of data types, leading to faster or
less memory-consuming implementations \cite{levillain.11.gretsi}.

\subsubsection{Acknowledgments} The authors thank Jacques-Olivier
Lachaud, who reviewed this paper, for his valuable comments, as well
the initial reviewers from the WADGMM workshop.

This work has been conducted in the context of the
SCRIBO project (\url{http://www.scribo.ws/}) of the Free Software
Thematic Group, part of the ``System@tic Paris-R\'egion'' Cluster
(France).  This project is partially funded by the French
Government, its economic development agencies, and by the
Paris-R\'egion institutions.

%
%

\bibliographystyle{splncs}
\bibliography{article,lrde,olena,sip}

\begin{thebibliography}{10}

\bibitem{dgtal.www.local}
{DGtal}:
\newblock Digital geometry tools and algorithms.
\newblock \url{http://liris.cnrs.fr/dgtal/}

\bibitem{geraud.08.mpool}
G\'eraud, {\relax Th}., Levillain, R.:
\newblock Semantics-driven genericity: A sequel to the static {C++}
  object-oriented programming paradigm ({SCOOP 2}).
\newblock In: Proceedings of the 6th International Workshop on Multiparadigm
  Programming with Object-Oriented Languages (MPOOL), Paphos, Cyprus (July
  2008)

\bibitem{levillain.09.ismm}
Levillain, R., G\'eraud, {\relax Th}., Najman, L.:
\newblock {Milena}: Write generic morphological algorithms once, run on many
  kinds of images.
\newblock In Wilkinson, M.H.F., Roerdink, J.B.T.M., eds.: Mathematical
  Morphology and Its Application to Signal and Image Processing -- Proceedings
  of the Ninth International Symposium on Mathematical Morphology (ISMM).
  Volume 5720 of Lecture Notes in Computer Science., Groningen, The
  Netherlands, Springer Berlin / Heidelberg (August 2009)  295--306

\bibitem{dornellas.03.ijprai}
d'Ornellas, M.C., van~den Boomgaard, R.:
\newblock The state of art and future development of morphological software
  towards generic algorithms.
\newblock International Journal of Pattern Recognition and Artificial
  Intelligence \textbf{17}(2) (March 2003)  231---255

\bibitem{olena.www}
{EPITA Research and Developpement Laboratory (LRDE)}:
\newblock The {Olena} image processing platform.
\newblock \url{http://olena.lrde.epita.fr}

\bibitem{lamy.07.cmpb}
Lamy, J.:
\newblock Integrating digital topology in image-processing libraries.
\newblock Computer Methods and Programs in Biomedicine \textbf{85}(1) (2007)
  51--58

\bibitem{ibanez.05.book}
Ib\'a{\~n}ez, L., Schroeder, W., Ng, L., Cates, J., the Insight
  Software~Consortium:
\newblock The {ITK} Software Guide. second edn.
\newblock Kitware, Inc. (November 2005)

\bibitem{itk.www}
{National Library of Medicine}:
\newblock Insight segmentation and registration toolkit ({ITK}).
\newblock \url{http://www.itk.org/}

\bibitem{levillain.10.icip}
Levillain, R., G\'eraud, {\relax Th}., Najman, L.:
\newblock Why and how to design a generic and efficient image processing
  framework: The case of the {Milena} library.
\newblock In: Proceedings of the IEEE International Conference on Image
  Processing (ICIP), Hong Kong (September 2010)  1941--1944

\bibitem{bertrand.07.chapter}
Bertrand, G., Couprie, M.:
\newblock Transformations topologiques discr\`etes.
\newblock In Coeurjolly, D., Montanvert, A., Chassery, J.M., eds.:
  G\'eom\'etrie discr\`ete et images num\'eriques.
\newblock Hermes Sciences Publications (2007)  187--209

\bibitem{couprie.09.pami}
Couprie, M., Bertrand, G.:
\newblock New characterizations of simple points in {2D}, {3D}, and {4D}
  discrete spaces.
\newblock IEEE Transactions on Pattern Analysis and Machine Intelligence
  \textbf{31}(4) (April 2009)  637--648

\bibitem{cousty.09.iwcia}
Cousty, J., Bertrand, G., Couprie, M., Najman, L.:
\newblock Collapses and watersheds in pseudomanifolds.
\newblock In: Proceedings of the 13th International Workshop on Combinatorial
  Image Analysis (IWCIA), Springer-Verlag (2009)  397--410

\bibitem{couprie.01.jei}
Couprie, M., Bezerra, F.N., Bertrand, G.:
\newblock Topological operators for grayscale image processing.
\newblock Journal of Electronic Imaging \textbf{10}(4) (2001)  1003--1015

\bibitem{dornellas.01.phd}
d'Ornellas, M.C.:
\newblock Algorithmic Patterns for Morphological Image Processing.
\newblock PhD thesis, Universiteit van Amsterdam (2001)

\bibitem{levillain.11.gretsi}
Levillain, R., G\'eraud, {\relax Th}., Najman, L.:
\newblock Une approche g\'en\'erique du logiciel pour le traitement d'images
  pr\'eservant les performances.
\newblock In: Proceedings of the 23rd Symposium on Signal and Image Processing
  (GRETSI), Bordeaux, France (September 2011) In French.

\end{thebibliography}

\end{document}